\documentclass[prb,showpacs,showkeys,preprint]{revtex4}

\usepackage{graphicx}

\begin{document}

\title{Multi-agent systems, Equiprobability, 
Gamma distributions and other Geometrical questions}

\author{Ricardo L\'opez-Ruiz}
\email{rilopez@unizar.es}
\affiliation{
DIIS and BIFI, Facultad de Ciencias, \\
Universidad de Zaragoza, E-50009 Zaragoza, Spain}

\author{Jaime Sa\~nudo }
\email{jsr@unex.es}
\affiliation{
Departamento de F\'isica, Facultad de Ciencias, \\
Universidad de Extremadura, E-06071 Badajoz, Spain}

\author{Xavier Calbet}
\email{xcalbet@googlemail.es}
\affiliation{
Instituto de Astrof{\'\i}sica de Canarias, \\
V{\'\i}a L\'actea, s/n, 
E-38200 La Laguna, Tenerife, Spain}

\date{\today}

\begin{abstract}

A set of many identical interacting agents obeying a global additive constraint is considered. 
Under the hypothesis of equiprobability in the high-dimensional volume 
delimited in phase space by the constraint, the statistical behavior of a generic agent
over the ensemble is worked out. The asymptotic distribution of that statistical behavior
is derived from geometrical arguments.
This distribution is related with the Gamma distributions found in several multi-agent 
economy models. The parallelism with all these systems is established.  
Also, as a collateral result, a formula for the volume of high-dimensional symmetrical 
bodies is proposed. 

\end{abstract}


\maketitle

\section{Introduction}

In this paper, a general multi-agent system under an additive constraint
is considered. Following the derivations done in 
Refs. \onlinecite{lopez2007-1,lopez2007-2,lopez2007-3},
we work out the geometrical properties of this system in phase space
in order to obtain its statistical behavior.
We observe a striking coincidence. 
The different dynamical mechanisms that have been proposed 
in the literature to model the interaction among agents in multi-agent 
economic systems \cite{yakovenko1,chakraborti2000,patriarca2004,angle2006,patriarca2006} 
display the same statistical results than those derived from
the geometrical properties of our system. This fact seems to suggest a close relationship
between the local interactions among the agents in the former systems and 
the global geometrical conformation in phase space of our general system.
We are inclined to think that all those dynamical mechanisms provoke that
those systems, with an adequate change of coordinates, evolve equiprobably 
over the volume of accessible states in the  transformed phase space.

We start in Section II by recalling the particular results obtained in Ref. \onlinecite{lopez2007-3}
for the cases in which the constraint has a linear or quadratic dependence on the variables.
Then, in Section III, we obtain the statistical behavior for a more general 
constraint. In Section IV, we establish a possible
relationship with other systems \cite{yakovenko1,chakraborti2000,patriarca2004,angle2006,patriarca2006} 
in which the Gamma distributions are also obtained. A formula
for the volume of high-dimensional symmetrical bodies is proposed in Section V.
The last Section VI contains our conclusions.

\section{Recalling some results} 

{\bf (A)} Let us assume $N$ agents interacting in an open economy \cite{lopez2007-3}, 
each one with coordinate $x_i$, $i=1,\ldots,N$, 
with $x_i\geq 0$ representing the wealth or money of the agent $i$,
and a total available amount of money $E$. The additive constraint reads:
\begin{equation}
x_1+x_2+\cdots +x_{N-1}+x_N \leq E.
\label{eq-e}
\end{equation} 
Under random evolution rules for the exchanging of money among agents \cite{yakovenko1},
let us suppose that this system evolves in the interior of the $N$-dimensional pyramid 
given by Eq. (\ref{eq-e}). We can suppose that
the state or the bank system of western societies plays in this model
the role of a heat reservoir that supplies money instead of energy.
The formula for the volume $V_N(E)$ of an equilateral $N$-dimensional pyramid 
formed by $N+1$ vertices linked by $N$ perpendicular sides of length $E$ is
\begin{equation}
V_N(E) = {E^N\over N!}.
\label{eq-S_n1}
\end{equation}
If each point on the $N$-dimensional pyramid is equiprobable, 
then the probability $f(x_i)dx_i$ of finding 
the agent $i$ with money $x_i$, with normalization condition
$\int_{0}^Ef(x_i)dx_i = 1$, is proportional to the 
volume formed by all the points into the $(N-1)$-dimensional pyramid 
having the $i$th-coordinate equal to $x_i$. 
We have shown \cite{lopez2007-3} that $f(x_i)$ verifies 
\begin{equation}
f(x_i) = {V_{N-1}(E-x_i)\over V_N(E)}.
\label{eq-volume1}
\end{equation}
If we call $\epsilon$ the mean wealth per agent, 
$E=N\epsilon$, then in the limit of large $N$ ($N\rightarrow\infty$), we have
\begin{equation}
\lim_{N\gg 1}{V_{N-1}(E-x)\over V_N(E)}={1\over \epsilon}\,e^{-{x/\epsilon}}dx,
\label{eq-gauss111}
\end{equation}
where the index $i$ has been removed because the distribution is the same for each agent,
and thus the wealth distribution can be obtained by averaging over all the agents,
\begin{equation}
f(x) =\epsilon^{-1}\,e^{-{x/\epsilon}}dx.
\label{eq-exp1}
\end{equation} 
This Boltzmann-Gibbs distribution has been found to fit the real distribution of incomes 
in western societies\cite{yakovenko1}.

{\bf (B)}
Now let us suppose a one-dimensional ideal gas of $N$ non-identical 
classical particles with masses $m_i$, with $i=1,\ldots,N$, and total 
maximum energy $E$. If particle
$i$ has a momentum $m_iv_i$, we define a kinetic energy:
\begin{equation}
K \equiv p_i^2 \equiv {1 \over 2}{ m_iv_i^2},
\label{eq-p_i}
\end{equation} 
where $p_i$ is the square root of the kinetic energy. 
Then the constraint reads: 
\begin{equation}
p_1^2+p_2^2+\cdots +p_{N-1}^2+p_N^2 \leq E.
\label{eq-E}
\end{equation} 
We see that the system has accessible states with different energy, which is 
supposed to be supplied by a heat reservoir. 
These states are all those enclosed into the volume 
of the $N$-sphere given by Eq. (\ref{eq-E}), with radius $E^{1/2}$. 
The formula for the volume $V_N(R)$
of an $N$-sphere of radius $R$ is
\begin{equation}
V_N(R) = {\pi^{N\over 2}\over \Gamma({N\over 2}+1)}R^{N},
\label{eq-S_n}
\end{equation}
where $\Gamma(\cdot)$ is the Gamma function. If we suppose that each point
into the $N$-sphere is equiprobable, then the probability $f(p_i)dp_i$ of finding 
the particle $i$, with coordinate $p_i$ (energy $p_i^2$) and normalization condition
$\int_{-R}^Rf(p_i)dp_i = 1$, is proportional to the 
volume formed by all the points into the $(N-1)$-sphere having the $i$th-coordinate 
equal to $p_i$. We have shown\cite{lopez2007-3} that $f(p_i)$ verifies
\begin{equation}
f(p_i) = {V_{N-1}((E-p_i^2)^{1/2})\over V_N(E^{1/2})}.
\label{eq-volume2}
\end{equation}

If we call $\epsilon$ the mean energy per particle, 
$E=N\epsilon$, then in the limit of large $N$ ($N\rightarrow\infty$), we have
\begin{equation}
\lim_{N\gg 1}{V_{N-1}((E-p^2)^{1/2})\over V_N(E^{1/2})}=
\sqrt{1\over 2\pi}\,\epsilon^{-1/2}\,e^{-{p^2/2\epsilon}},
\label{eq-gauss2}
\end{equation}
where the index $i$ has been removed because the distribution is the same for each particle. 
Thus the asymptotic distribution 
\begin{equation}
f(p)=\sqrt{1\over 2\pi}\,\epsilon^{-1/2}\,e^{-{p^2/2\epsilon}}
\label{eq-gauss22}
\end{equation}
can be obtained by averaging over all the particles.
If the change of variables $p=\sqrt{m\over 2}\,v$
is performed,
with $v$ the generic velocity of a particle,
then the Maxwellian distribution is just derived from geometrical arguments.

\section{Multi-agent systems and equiprobability:\\
general derivation of the asymptotic distribution}

In this section, we address the same problem above presented but in a general way.
Let $b$ be a positive real constant (cases $b=1,2$ have been indicated in the former section). 
If we have a set of positive variables $(x_1,x_2,\ldots,x_N)$ verifying 
\begin{equation}
x_1^b+x_2^b+\cdots +x_{N-1}^b+x_N^b \leq E
\label{eq-Ek}
\end{equation}
with an adequate mechanism assuring 
the equiprobability of all the possible states $(x_1,x_2,\ldots,x_N)$
into the volume given by expression (\ref{eq-Ek}),
will we have for the generic variable $x$ the distribution
\begin{equation}
f(x)dx \sim \epsilon^{-1/b}\,e^{-{x^b/b\epsilon}}dx,
\label{eq-gaussn}
\end{equation}
when we average over the ensemble in the limit $N,E\rightarrow\infty$,
with $E=N\epsilon$, and constant $\epsilon$?. 
Now we show that the answer is affirmative. 

From the cases $b=1,2$, (see Eqs. (\ref{eq-volume1}) and (\ref{eq-volume2})),
we can extrapolate the general formula that will give
us the statistical behavior $f(x)$ of the generic variable $x$, 
when the system runs equiprobably into the volume defined by a constraint of type (\ref{eq-Ek}).
The probability $f(x)dx$ of finding 
an agent with generic coordinate $x$ is proportional to the 
volume $V_{N-1}((E-x^b)^{1/b})$ formed by all the points into the $(N-1)$-dimensional 
symmetrical body limited by the constraint $(E-x^b)$. 
Thus, the $N$-dimensional volume can be written as
\begin{equation}
V_N(E^{1/b})= \int_{0}^{E^{1/b}}\,V_{N-1}((E-x^b)^{1/b})\,dx.
\label{eq-volumen}
\end{equation}
Taking into account the normalization condition $\int_{0}^{E^{1/b}}f(x)dx = 1$,
the expression for $f(x)$ is obtained:
\begin{equation}
f(x) = {V_{N-1}((E-x^b)^{1/b})\over V_N(E^{1/b})}.
\label{eq-volume3}
\end{equation}

The $N$-dimensional volume, $V_N(b,\rho)$, of a $b$-symmetrical body with side of length $\rho$ is 
proportional to the term $\rho^N$ and to a coefficient $g_b(N)$ that depends on $N$:
\begin{equation}
V_N(b,\rho)=g_b(N)\,\rho^N.
\label{eq-volumenn}
\end{equation}
The parameter $b$ indicates the original equation (\ref{eq-Ek}) that defines the 
boundaries of the volume $V_N(b,\rho)$. Thus, for instance, from Eq. (\ref{eq-S_n1}), 
we have $g_{b=1}(N) =  1/ N!$.

Coming back to Eq. (\ref{eq-volume3}), we can manipulate $V_{N}((E-x^b)^{1/b})$
to obtain (the index $b$ is omitted in the formule of $V_N$):
\begin{equation}
V_N((E-x^b)^{1/b})=g_b(N)\,\left[(E-x^b)^{1/b}\right]^N=g_b(N)\,E^{N\over b}\,
\left(1-{x^b\over E}\right)^{N\over b}.
\label{eq-volumexb}
\end{equation}
If we suppose $E=N\epsilon$, then $\epsilon$ represents the mean value of $x^b$
in the collectivity, that is, $\epsilon=<x^b>$. If $N$ tends toward infinity,
it results:
\begin{equation}
\lim_{N\gg 1}\left(1-{x^b\over E}\right)^{N\over b} \, =\, e^{-x^b/ b\epsilon}.
\label{eq-volumexb1}
\end{equation}
Thus,
\begin{equation}
V_N((E-x^b)^{1/b})=V_N(E^{1/b})\,e^{-x^b/ b\epsilon}.
\label{eq-volumexb2}
\end{equation}
Substituting this last expression in formula (\ref{eq-volume3}),
the exact form for $f(x)$ is found in the thermodynamic limit
($N,E\rightarrow\infty$):
\begin{equation}
f(x)dx = c_b \,\epsilon^{-1/b}\,e^{-{x^b/b\epsilon}}dx,
\label{eq-fx}
\end{equation}
with $c_b$ given by
\begin{equation}
c_b\,=\,{g_b(N-1)\over g_b(N)\,N^{1/b}}.
\label{eq-cb}
\end{equation}
Hence, the conjecture (\ref{eq-gaussn}) is proved.

Doing a thermodinamical simile, 
we can calculate the dependence of $\epsilon$ on the temperature 
by differentiating the entropy with respect to the energy. 
The entropy can be written as $S=-kN\!\int_{0} 
^{\infty} f(x)\ln f(x)\,dx$, where $f(x)$ is given by Eq.~(\ref{eq-fx})
and $k$ is the Boltzmann constant. 
If we recall that $\epsilon=E/N$, we obtain
\begin{equation}
S(E)= {kN\over b}\ln\left({E\over N} \right) + {kN\over b}(1-b\ln c_b),
\end{equation}
where it has been used that $\epsilon=<x^b>=\!\int_{0} ^{\infty} x^bf(x)dx$.

The calculation of the temperature $T$ gives
\begin{equation}
T^{-1}= \left({\partial S\over \partial E} \right)_N = {kN\over bE} = {k\over b\epsilon}.
\end{equation}
Thus $\epsilon=kT/b$, a result that recovers the theorem of equipartition of energy
for the quadratic case $b=2$.
The distribution for all $b$ is finally obtained:
\begin{equation}
f(x)dx = c_b\left({b\over kT}\right)^{1/b}\,e^{-x^b/kT}dx.
\end{equation}

\section{Gamma distributions}

If we perform the change of variables $y=\epsilon^{-1/b}x$ in the normalization
condition of $f(x)$, $\!\int_{0} ^{\infty} f(x)dx=1$, we find that
\begin{equation}
c_b=\left[\!\int_{0} ^{\infty} e^{-y^b/b}\,dy\right]^{-1}.
\label{eq-cb1}
\end{equation}
If we introduce the new variable $z=y^b/b$, the distribution $f(x)$ as function of $z$ reads:
\begin{equation}
f(z)dz = {c_b\over b^{1-{1\over b}}} \,z^{{1\over b}-1}\,e^{-z}\,dz.
\label{eq-fz}
\end{equation}
Let us observe that the Gamma function appears in the normalization condition,
\begin{equation}
\int_{0} ^{\infty} f(z)dz={c_b\over b^{1-{1\over b}}} \,\int_{0} ^{\infty}
\,z^{{1\over b}-1}\,e^{-z}\,dz = {c_b\over b^{1-{1\over b}}}\,\Gamma\left({1\over b}\right)=1.
\label{eq-fz1}
\end{equation}
This implies that
\begin{equation}
c_b={b^{1-{1\over b}} \over \Gamma\left({1\over b}\right)}.
\label{eq-cb2}
\end{equation}
By using Mathematica the positive constant $c_b$ is plotted versus $b$ in Fig. 1.
We see that $\lim_{b\rightarrow 0}c_b=\infty$, and that $\lim_{b\rightarrow \infty}c_b=1$.
The minimum of $c_b$ is reached for $b=3.1605$, taking the value $c_b=0.7762$.
Still further, we can calculate from Eq. (\ref{eq-cb2})
the asymptotic dependence of $c_b$ on b: 
\begin{eqnarray}
\lim_{b\rightarrow 0}c_b & = & \sqrt{1\over 2\pi}\,\sqrt{b}\,e^{1/b}\left(1-{b\over 12}+\cdots\right), \label{eq-cb3}\\
&& \nonumber\\
\lim_{b\rightarrow \infty}c_b & = & b^{-1/b}\left(1 + {\gamma\over b}+\cdots\right),\label{eq-cb4}
\end{eqnarray}
where $\gamma$ is the Euler constant, $\gamma=0.5772$. The asymptotic function (\ref{eq-cb3}) is
obtained after substituting in (\ref{eq-cb2}) the value of $\Gamma(1/b)$ by $(1/b-1)!$, and performing the
Stirling approximation on this last expression, knowing that $1/b\rightarrow\infty$. The function
(\ref{eq-cb4}) is found after looking for the first Taylor expansion terms of the Gamma function
around the origin $x=0$. They can be derived from the Euler's reflection formula, 
$\Gamma(x)\Gamma(1-x)=\pi/\sin(\pi x)$. We obtain $\Gamma(x\rightarrow 0)=x^{-1}+\Gamma'(1)+\cdots$.
From here, recalling that $\Gamma'(1)=-\gamma$, we get $\Gamma(1/b)=b-\gamma+\cdots$, 
when $b\rightarrow\infty$. Although this last term of the Taylor expansion, $-\gamma$, 
is negligible we maintain it in expression (\ref{eq-cb4}). The only minimum of $c_b$ is reached for
the solution $b=3.1605$ of the equation $\psi(1/b)+\log b+b-1=0$, where $\psi(\cdot)$ is the 
digamma function (see Fig. 1).

Let us now recall two interesting statistical economic models that display a statistical behavior
given by distributions of the form (\ref{eq-fz}), that is, the standard Gamma distributions 
with shape parameter $1/b$,

\begin{equation}
f(z)dz = {1\over \Gamma({1\over b})} \,z^{{1\over b}-1}\,e^{-z}\,dz.
\label{eq-fg}
\end{equation}
\vspace {0.5 cm}

{\bf MODEL A:} The first one is the saving propensity model introduced by 
Chakraborti and Chakrabarti \cite{chakraborti2000}. In this model a set of $N$ economic
agents, having each agent $i$ (with $i=1,2,\cdots,N$) an amount of money, $u_i$,
exchanges it under random binary $(i,j)$ interactions, $(u_i,u_j)\rightarrow (u_i',u_j')$, 
by the following the exchange rule:
\begin{eqnarray}
u'_i & = & \lambda u_i+\epsilon(1-\lambda)(u_i+u_j), \\
u'_j & = & \lambda u_j+\bar\epsilon(1-\lambda)(u_i+u_j), 
\end{eqnarray}
with $\bar\epsilon=(1-\epsilon)$, and $\epsilon$ a random number in the interval $(0,1)$.
The parameter $\lambda$, with $0<\lambda<1$, is fixed, and represents the fraction of money
saved before carrying out the transaction. Let us observe that money is conserved, i.e.,
$u_i+u_j=u_i'+u_j'$, hence in this model the economy is closed. Defining the parameter
$n(\lambda)$ as
\begin{equation}
n(\lambda)={1+2\lambda \over 1-\lambda},
\label{eq-nl1}
\end{equation}
and scaling the wealth of the agents as $\bar z=nu/<u>$, with $<u>$ representing the average money
over the ensemble of agents, it is found that the asymptotic wealth distribution in this system
obeys the standard Gamma distribution\cite{patriarca2004} 
\begin{equation}
f(\bar z)d\bar z = {1\over \Gamma(n)} \,\bar z^{n-1}\,e^{-\bar z}\,d\bar z.
\label{eq-fg1}
\end{equation}
The case $n=1$, which means a null saving propensity, $\lambda=0$, recovers the model
of Dragulescu and Yakovenko \cite{yakovenko1} in which the Gibbs distribution is observed.
If  we compare Eqs. (\ref{eq-fg1}) and (\ref{eq-fg}), a close relationship between this
economic model and the geometrical problem solved in the former section can be established. 
It is enough to make 
\begin{eqnarray}
n & = & 1/b, \label{eq-nb}\\
\bar z & = & z\label{eq-nz}, 
\end{eqnarray}
to have two equivalent systems. This means that, from Eq. (\ref{eq-nb}), 
we can calculate $b$ from the saving parameter $\lambda$ with the formula
\begin{equation}
b={1-\lambda \over 1+2\lambda}.
\label{eq-nl11}
\end{equation}
As $\lambda$ takes its values in the interval $(0,1)$,
then the parameter $b$ also runs in the same interval $(0,1)$. 
On the other hand, recalling that $z=x^b/b\epsilon$,
we can get the equivalent variable $x$ from Eq. (\ref{eq-nz}),
\begin{equation}
x=\left[{\epsilon\over <u>}\;u\;\right]^{1/b}, 
\label{eq-nl12}
\end{equation}
where $\epsilon$ is a free parameter that determines the mean value
of $x^b$ in the equivalent geometrical system. 
Formula (\ref{eq-nl12}) means to perform the change of variables 
$u_i\rightarrow x_i$, with $i=1,2,\cdots,N$, for all the particles/agents of the ensemble.
Then, we conjecture that the economic system represented by the generic 
pair $(\lambda,u)$, when it is transformed in
the geometrical system given by the generic pair $(b,x)$,
as indicated by the rules (\ref{eq-nl11}) and (\ref{eq-nl12}),
runs in an equiprobable form on the surface defined by the 
relationship (\ref{eq-Ek}), where the inequality has been transformed
in equality. This last detail is due to the fact the economic system is closed,
and then it conserves the total money, whose equivalent quantity in the geometrical
problem is $E$. If the economic system were open, with an upper limit in the wealth, 
then the transformed system would evolve in an equiprobable way over the volume
defined by the inequality (\ref{eq-Ek}). To see more clearly the equivalence 
between surface and volume in a statistical ensemble we address the reader 
to Ref. \onlinecite{lopez2007-3} where this question has been discussed for similar cases.

\vspace {0.5 cm}
{\bf MODEL B:} 
The second one is the model introduced by Angle \cite{angle2006}. 
In this model a set of $N$ economic
agents, having each agent $i$ (with $i=1,2,\cdots,N$) an amount of money, $u_i$,
exchanges it under random binary $(i,j)$ interactions, $(u_i,u_j)\rightarrow (u_i',u_j')$, 
by the following the exchange rule:
\begin{eqnarray}
u'_i & = & u_i-\Delta u, \\
u'_j & = & u_j+\Delta u, 
\end{eqnarray}
where 
\begin{equation}
\Delta u=\eta(x_i-x_j)\,\epsilon\omega x_i-[1-\eta(x_i-x_j)]\,\epsilon\omega x_j,
\label{eq-delta}
\end{equation}
with $\epsilon$ a random number in the interval $(0,1)$. The exchange parameter, $\omega$, 
represents the maximum fraction of wealth lost by one of the two interacting agents ($0<\omega< 1$).
Whether the agent who is going to loose part of the money is the $i$-th or the $j$-th agent,
depends nonlinearly on $(x_i-x_j)$, and this is decided by the random dichotomous function $\eta(t)$:
$\eta(t>0)=1$ (with additional probability $1/2$) and $\eta(t<0)=0$ (with additional probability $1/2$). 
Hence, when $x_i>x_j$,  the value $\eta=1$ produces a wealth transfer from agent $i$ to agent $j$
with probability $1/2$, and when $x_i<x_j$,  the value $\eta=0$ produces a wealth transfer from 
agent $j$ to agent $i$ with probability $1/2$.
Defining in this case the parameter $n(\omega)$ as
\begin{equation}
n(\omega)={3-2\omega \over 2\omega},
\label{eq-na1}
\end{equation}
and scaling the wealth of the agents as $\bar z=nu/<u>$, with $<u>$ representing the average money
over the ensemble of agents, it is found that the asymptotic wealth distribution in this system
obeys the standard Gamma distribution\cite{patriarca2006} 
\begin{equation}
f(\bar z)d\bar z = {1\over \Gamma(n)} \,\bar z^{n-1}\,e^{-\bar z}\,d\bar z.
\label{eq-fg11}
\end{equation}
The case $n=1$, which means an exchange parameter $\omega=3/4$, recovers the model
of Dragulescu and Yakovenko \cite{yakovenko1} in which the Gibbs distribution is observed.
If  we compare Eqs. (\ref{eq-fg11}) and (\ref{eq-fg}), a close relationship between this
economic model and the geometrical problem solved in the last section can be established. 
It is enough to make 
\begin{eqnarray}
n & = & 1/b, \label{eq-nb1}\\
\bar z & = & z\label{eq-nz1}, 
\end{eqnarray}
to have two equivalent systems. This means that, from Eq. (\ref{eq-nb1}), 
we can calculate $b$ from the exchange parameter $\omega$ with the formula
\begin{equation}
b={2\omega \over 3-2\omega}.
\label{eq-nl22}
\end{equation}
As $\omega$ takes its values in the interval $(0,1)$,
then the parameter $b$ runs in the interval $(0,2)$. 
It is curious to observe that in this model the interval $\omega\in(3/4,1)$ maps on $b\in(1,2)$,
a fact that does not occur in MODEL A.
On the other hand, recalling that $z=x^b/b\epsilon$,
we can get the equivalent variable $x$ from Eq. (\ref{eq-nz1}),
\begin{equation}
x=\left[{\epsilon\over <u>}\;u\;\right]^{1/b}.
\label{eq-nl23}
\end{equation}
where $\epsilon$ is a free parameter that determines the mean value
of $x^b$ in the equivalent geometrical system. 
Formula (\ref{eq-nl23}) means to perform the change of variables 
$u_i\rightarrow x_i$, with $i=1,2,\cdots,N$, for all the particles/agents of the ensemble.
Then, we conjecture that the economic system represented by the generic 
pair $(\lambda,u)$, when it is transformed in
the geometrical system given by the generic pair $(b,x)$,
as indicated by the rules (\ref{eq-nl22}) and (\ref{eq-nl23}),
runs in an equiprobable form on the surface defined by the 
relationship (\ref{eq-Ek}), where the inequality has been transformed
in equality. As explained above, 
this last detail is due to the fact the economic system is closed,
and then it conserves the total money, whose equivalent quantity in the geometrical
problem is $E$. If the economic system were open, with an upper limit in the wealth, 
then the transformed system would evolve in an equiprobable way over the volume
defined by the inequality (\ref{eq-Ek}). This equivalence in using 
the surface or the volume of a statistical ensemble in order to obtain its statistical
behavior has been discussed in Ref. \onlinecite{lopez2007-3} for similar cases.

\section{Other geometrical questions}

We shall proceed now to derive an asymptotic formula ($N\rightarrow\infty$) for the volume of
the $N$-dimensional symmetrical body enclosed by the surface 
\begin{equation}
x_1^b+x_2^b+\cdots +x_{N-1}^b+x_N^b = E.
\label{eq-Ekk}
\end{equation}
The linear dimension $\rho$ of this volume, i.e., the length of one of its sides verifies
$\rho\sim E^{1/b}$. 
As argued in Eq. (\ref{eq-volumenn}),
the $N$-dimensional volume, $V_N(b,\rho)$, is 
proportional to the term $\rho^N$ and to a coefficient $g_b(N)$ 
that depends on $N$. Thus,
\begin{equation}
V_N(b,\rho)=g_b(N)\,\rho^N,
\label{eq-volumenn1}
\end{equation}
where the characteristic $b$ indicates the particular boundary given by equation (\ref{eq-Ekk}).

For instance, from Eq. (\ref{eq-S_n1}), we can write in a formal way:
\begin{equation}
g_{b=1}(N) =  {1^{N\over 1}\over \Gamma({N\over 1}+1)}.
\label{eq-gN}
\end{equation}
From Eq. (\ref{eq-S_n}), if we take the diameter, $\rho=2R$, as the linear dimension of the $N$-sphere,
we obtain:
\begin{equation}
g_{b=2}(N) =  {\left({\pi\over 4}\right)^{N\over 2}\over \Gamma\left({N\over 2}+1\right)}.
\label{eq-gN1}
\end{equation}
These expressions (\ref{eq-gN}) and (\ref{eq-gN1}) suggest a possible general formula
for the factor $g_{b}(N)$, let us say 
\begin{equation}
g_{b}(N) =  {a^{{N\over b}}\over \Gamma\left({N\over b}+1\right)},
\label{eq-gN2}
\end{equation}
where $a$ is a $b$-dependent constant to be determined. For example,
$a=1$ for $b=1$ and $a=\pi/4$ for $b=2$. 

In order to find the dependence of $a$ on the parameter $b$, the regime $N\rightarrow \infty$ 
is supposed. Applying Stirling approximation for the factorial $({N\over b})!$ in the denominator
of expression (\ref{eq-gN2}), and inserting it in expression (\ref{eq-cb}), 
it is straightforward to find out the relationship:
\begin{equation}
c_b \,= \,(ab)^{-1/b}.
\label{eq-gN3}
\end{equation}
From here and formula (\ref{eq-cb2}), we get:
\begin{equation}
a\,= \,\left[\Gamma\left({1\over b}+1\right)\right]^b,
\label{eq-gN4}
\end{equation}
that recovers the exact results for $b=1,2$. The behavior of $a$ 
is monotonous decreasing when $b$ is varied from $b=0$, 
where $a$ diverges as $a\sim 1/b+\cdots$, 
up to the limit $b\rightarrow\infty$, where $a$ decays asymptotically 
toward the value $a_{\infty}=e^{-\gamma}=0.5614$.
 
Hence, the formula for $g_{b}(N)$ is obtained:
\begin{equation}
g_{b}(N)\,= \,{\,\,\Gamma\left({1\over b}+1\right)^N\over \Gamma\left({N\over b}+1\right)},
\label{eq-gN5}
\end{equation}
It would be also possible to multiply this last expression (\ref{eq-gN5})
by a general polynomial $K(N)$ in the variable $N$,
and all the derivation done from Eq. (\ref{eq-gN2})
would continue to be correct. We omit this possibility
in our calculations. For a fixed $N$, we have that $g_b(N)$ increases monotonously
from $g_b(N)=0$, for $b=0$, up to $g_b(N)=1$, 
in the limit $b\rightarrow\infty$ (see Fig. 2).
For a fixed $b$, we have that $g_b(N)$ decreases monotonously
from $g_b(N)=1$, for $N=1$, up to $g_b(N)=0$, 
in the limit $N\rightarrow\infty$ (see Fig. 3).

The final result for the volume of an $N$-dimensional symmetrical body of
characteristic $b$ given by the boundary (\ref{eq-Ekk}) reads:
\begin{equation}
V_N(b,\rho)\,= \,{\,\,\Gamma\left({1\over b}+1\right)^N\over \Gamma\left({N\over b}+1\right)}\,\rho^N,
\label{eq-gN6}
\end{equation}
with $\rho\sim E^{1/b}$.

\section{Conclusions}

In this work, we have considered a general multi-agent open system verifying an
additive constraint. Its statistical behavior has been derived from geometrical
arguments. The Maxwellian and the Boltzmann-Gibbs distributions are particular cases
of this type of systems. Also, other multi-agent economy models, such as the Chakraborti and Chakrabarti's
model\cite{chakraborti2000}, the Angle's model \cite{angle2006}, 
and the Dragalescu and Yakovenko's model\cite{yakovenko1}, 
show similar statistical behaviors than our general system. This fact suggests a geometrical
interpretation of all those models. The equivalence with the Chakraborti and Chakrabarti's
model is established when the geometrical characteristic $b$ of our model runs in the interval $(0,1)$.
The equivalence with the Angle's model is established when $b$ varies in the interval $(0,2)$.
As a particular case of both types of model, the Dragulescu and Yakovenko's model is obtained for $b=1$.

Let us remark that we have not found in the literature other multi-agent models to establish 
an equivalence with our system in the range $b\in(2,\infty)$. This point remains as an open question 
and it can be a challenge that will probably trigger other works in this direction.

\begin{acknowledgements}
The authors acknowledge some financial support from Spanish DGICYT Projects FIS2005-06237
and and FIS2006-12781-C02-01.
\end{acknowledgements}


\newpage

$\;$\vspace{1cm}

\centerline{\bf Figures}
\vspace{1cm}

\begin{figure}[h]
\centerline{\includegraphics[width=9cm]{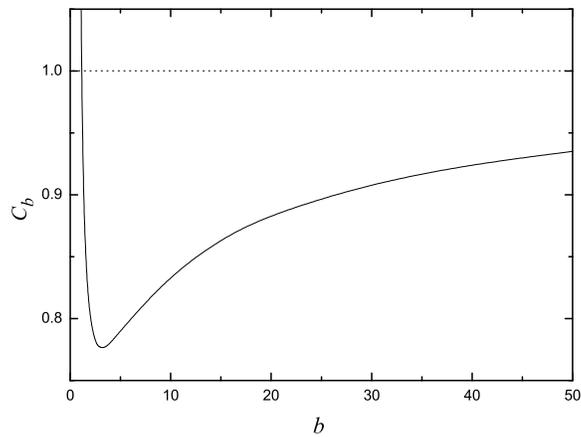}}
\caption{Normalization constant $c_b$ versus $b$, calculated from Eq. (\ref{eq-cb2}). 
The asymptotic behavior is: $\lim_{b\rightarrow 0}c_b=\infty$, and $\lim_{b\rightarrow \infty}c_b=1$.
This last asymptote is represented by the dotted line.
The minimum of $c_b$ is reached for $b=3.1605$, taking the value $c_b=0.7762$.} 
\label{fig1}
\end{figure}

\vspace{1cm}

\begin{figure}[h]
\centerline{\includegraphics[width=9cm]{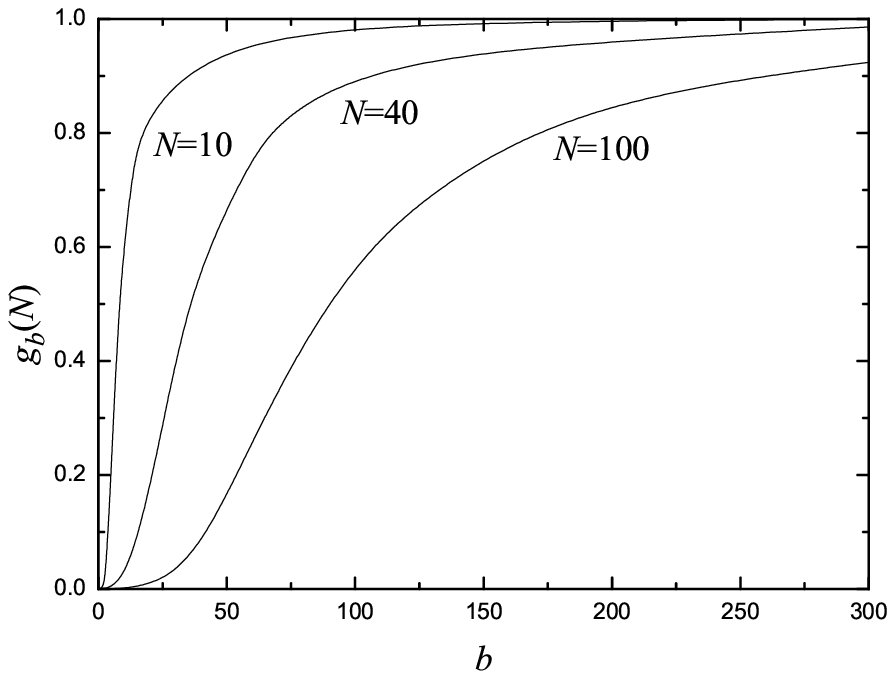}}
\caption{The factor $g_b(N)$ versus $b$ for $N=10, 40, 100$, calculated from Eq. (\ref{eq-gN5}). 
Observe that $g_b(N)=0$ for $b=0$,
and $\lim_{b\rightarrow \infty}g_b(N)=1$.} 
\label{fig2}
\end{figure}

\vspace{1cm}

\begin{figure}[h]
\centerline{\includegraphics[width=9cm]{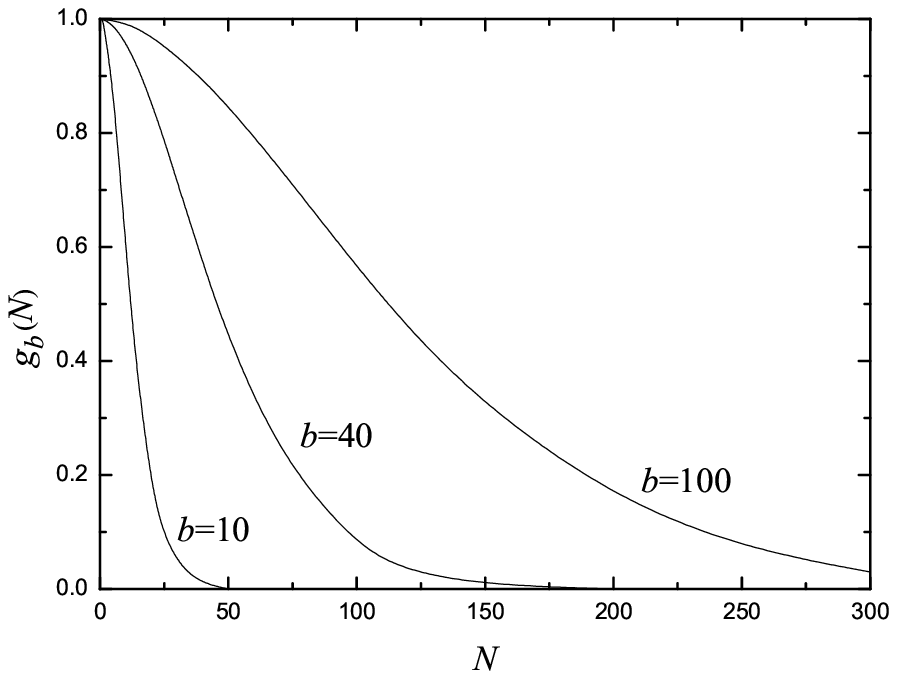}}
\caption{The factor $g_b(N)$ versus $N$ for $b=10, 40, 100$, calculated from Eq. (\ref{eq-gN5}). 
Observe that $g_b(N)=1$ for $N=1$,
and $\lim_{N\rightarrow \infty}g_b(N)=0$.} 
\label{fig3}
\end{figure}

\end{document}